\documentclass[12pt,twoside,fleqn]{article}
\usepackage{epsfig}
\usepackage{psfig}
\usepackage{dcolumn}
\usepackage[dvips]{rotating}
\def\lsim{\raise0.3ex\hbox{$<$\kern-0.75em\raise-1.1ex\hbox{$\sim$}}}
\def\gsim{\raise0.3ex\hbox{$>$\kern-0.75em\raise-1.1ex\hbox{$\sim$}}}
\makeatletter
\@addtoreset{equation}{section}
\makeatother

\newcommand{\beqn} {\begin{equation}}
\newcommand{\eqn} {\end{equation}}

\newcommand{\nn} {\nonumber}
\newcommand{\nuh} {\hat{\nu}}
\newcommand{\muh} {\hat{\mu}}
\newcommand{\pb} {\bar{\psi}}

\newcommand{\slsh}[1] {#1\kern-.43em/}
\newcommand{\real}{{\sf I}\kern-.12em{\sf R}}
\newcommand{\comp}{{\sf I}\kern-.48em{\sf C}}
\newcommand{\nin} {\in\kern-.6em/}

\newcommand{\ie}{{\sl i.e.~}}

\def\MEF{m_{\rm eff}}\def\mef{\ifmmode\MEF\else$\MEF$\fi}

\def\SM{s_{\mu}}\def\xm{\ifmmode\SM\else$\SM$\fi}

\newcommand{\re}{{\rm Re~}}
\newcommand{\tr}{{\rm Tr~}}
\newcommand{\plaq}{\mbox{\raisebox{-1.5mm}
{\epsfig{file=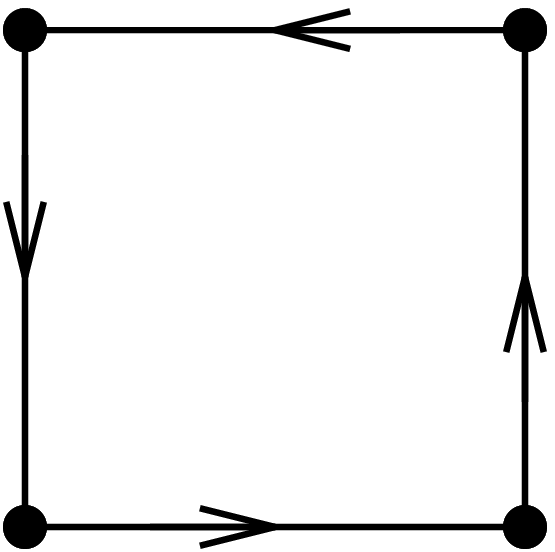,height=6mm
}}~}}
\newcommand{\loOp}{\mbox{\raisebox{-1.5mm}
{\epsfig{file=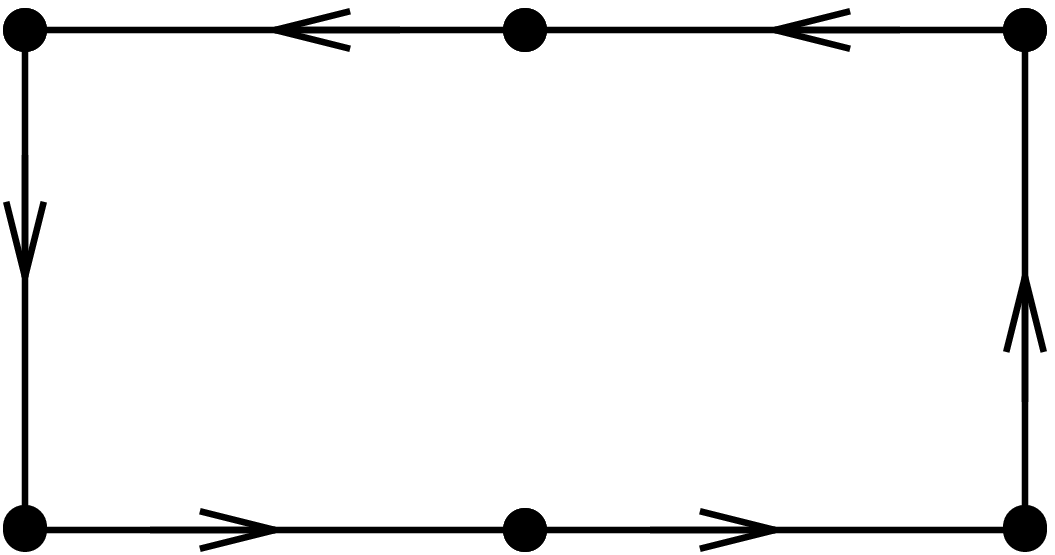,height=6mm
}}~}}
\newcommand{\lOop}{\mbox{\raisebox{-4mm}
{\epsfig{file=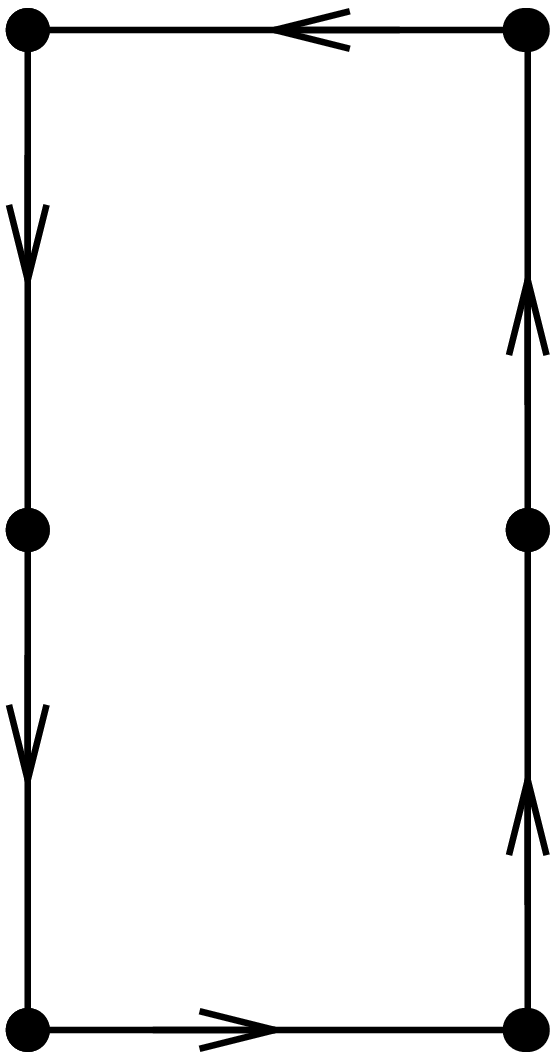,height=12mm
}}~}}
\newcommand{\fa}{\mbox{\raisebox{-0.6cm}
{\epsfig{file=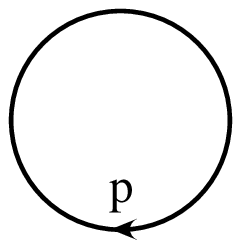,width=1.5cm
}}~}}
\newcommand{\ssa}{\mbox{\raisebox{-0.6cm}
{\epsfig{file=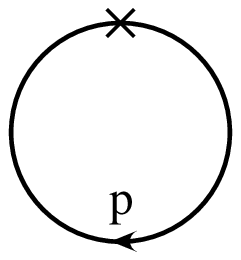,width=1.5cm
}}~}}
\newcommand{\ssb}{\mbox{\raisebox{-1.3cm}
{\epsfig{file=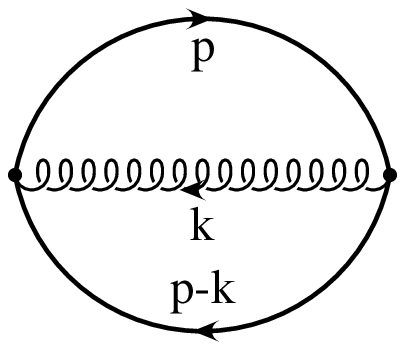,height=3cm
}}~}}
\newcommand{\ssc}{\mbox{\raisebox{-1.3cm}
{\epsfig{file=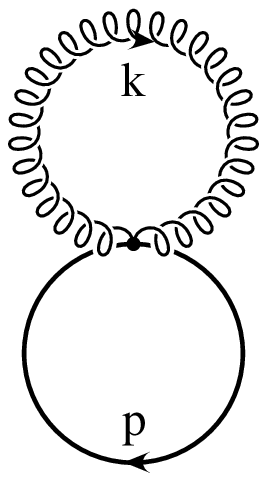,height=3cm
}}~}}

\newcommand{\propa}{\mbox{\raisebox{-7.0mm}
{\epsfig{file=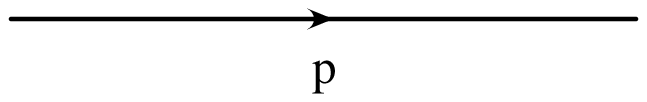, width=4cm
}}~}}
\newcommand{\propb}{\mbox{\raisebox{-7.0mm}
{\epsfig{file=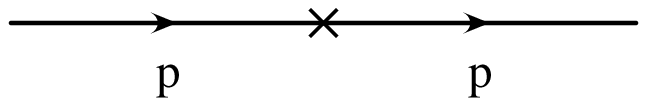, width=4cm
}}~}}
\newcommand{\sigmaa}{\mbox{\raisebox{-7mm}
{\epsfig{file=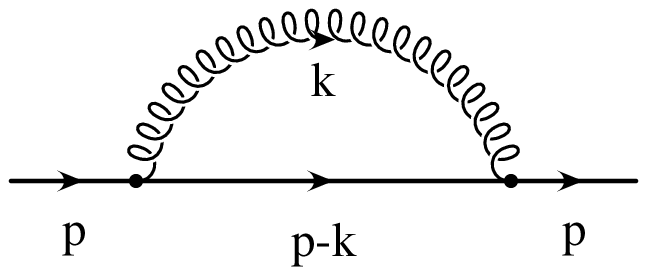,width=4cm
}}~}}
\newcommand{\sigmab}{\mbox{\raisebox{-7mm}
{\epsfig{file=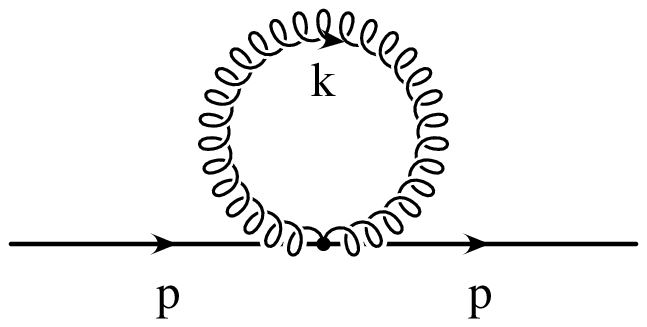,width=4cm
}}~}}
\begin{document}
\thispagestyle{empty}
%
\mbox{} \hfill BI-TP 99/1\\
\mbox{} \hfill FSU-SCRI-99-02\\
 \mbox{} \hfill January 1999\\
\begin{center}
{{\large \bf Improved Staggered Fermion Actions for \\
\vspace*{0.2cm}
QCD Thermodynamics}
\bigskip
} \\
\vspace*{1.0cm}
{\large U. M. Heller$^a$, F. Karsch$^b$ and B. Sturm$^b$} \\
\vspace*{1.0cm}
{\normalsize
$\mbox{}$ {$^a$
SCRI, Florida State University, Tallahassee, FL 32306-4130, USA}
}
\\

{\normalsize
$\mbox{}$ {$^b$ Fakult\"at f\"ur Physik, Universit\"at Bielefeld,
D-33615 Bielefeld, Germany}
}
\end{center}
\vspace*{1.0cm}
\centerline{\large ABSTRACT}

We analyze the cut-off dependence of the fermion contribution to
the finite temperature free energy density in ${\cal O}(g^2)$ lattice
perturbation theory for several improved staggered fermion actions.
Cut-off effects are drastically reduced for the Naik action and
an action with improved rotational symmetry of the quark propagator.
We show that improvement of rotational symmetry at ${\cal O}(g^2)$
further reduces cut-off effects in thermodynamic observables.
We also show that the introduction of fat-links does not have
a significant influence on cut-off distortions at ${\cal O}(g^2)$.

\baselineskip 20pt

\noindent

\vskip 20pt
\vfill
\eject
\baselineskip 15pt

\section{Introduction}

During recent years much progress has been made in the analysis of
the thermo\-dynamics of $SU(N)$ gauge theories. Based on detailed studies of
the cut-off dependence of thermodynamic observables in lattice regularized
gauge theories an extrapolation of results for the equation of state
to the continuum limit has been performed for the first time \cite{Boy95}.
In addition, it could be shown that the influence of a finite lattice
cut-off on thermodynamic observables is greatly reduced when improved
actions such as ${\cal O}(a^2)$ Symanzik improved \cite{Bei96} or tree
level perfect \cite{Pap96} gauge actions are used.  By construction tree
level improved actions do reduce the cut-off dependence of thermodynamic
observables in the infinite temperature, ideal gas limit of
gauge theories  \cite{Bei98,Kar98}. They do, however, also seem to
reduce cut-off effects at non-zero values of the gauge coupling $g^2$,
{\it i.e.} at finite temperature. Similar conclusions have been drawn from a
first analysis of QCD thermodynamics with an improved staggered fermion action \cite{Eng96}.

Using tree level improved gauge and fermion actions it is now possible
to perform lattice calculations at finite temperature, which at least
in the infinite temperature limit lead to acceptably small systematic
cut-off errors.
Numerical simulations for $SU(N)$ gauge theories as well as QCD
with dynamical fermions \cite{Eng96} give some indications that this
improvement carries over also into the temperature regime close
to the deconfinement phase transition. This suggests that improved
actions also show a reduced cut-off dependence at non-zero gauge coupling
$g^2$. It would, of course, be nice to go on with the improvement program
and construct systematically ${\cal O}(g^2 a^2)$ improved or even 1-loop
perfect actions which could be used for, e.g. thermodynamic, calculations.
However, it is well known that this generates a large number of new terms,
including 4-fermion operators, that would contribute to the 1-loop
improved fermion action and would
make such an action impractical for numerical calculations \cite{Luo}.
We thus will aim at a more moderate
goal in this paper. Within a class of tree level improved actions we will
look for actions which lead to small cut-off errors in thermodynamic
observables at 1-loop level.

In the case of the standard staggered fermion action it is known that
deviations from continuum perturbation theory are large for thermodynamic
observables also at ${\cal O}(g^2)$ \cite{Hel85}.
It is the purpose of this paper
to analyze in 1-loop lattice perturbation theory in how far
improved actions do reduce the cut-off dependence at this order.
Besides analyzing tree level improved gauge and fermion
actions we also discuss the influence of 1-loop improvement in the
gauge sector on fermionic observables and construct a new fermion action
with improved rotational symmetry at ${\cal O}(g^2)$.
Furthermore, we investigate in how far
the use of {\it fat links}, which have been introduced to improve the
flavour symmetry of the staggered fermion action \cite{fat}, does influence
the calculation of thermodynamic observables. Fat links modify
cut-off dependent terms in thermodynamic observables at ${\cal O}(g^2 a^2)$.

This paper is organized as follows. In the next section we introduce the
various staggered fermion actions we are going to analyze in 1-loop
perturbation theory and define the basic quantities needed in our perturbative
calculations. In the following subsections we  determine the coefficients of
these actions to improve rotational symmetry at tree-level and in 1-loop
order.
In section 3 we present results from a 1-loop calculation of the fermion
contribution to the free energy density at finite temperature.
In section 4 we give our conclusions. Some further details of our calculations are
summarized in the appendix.

\section{Improved fermion actions}

As starting point for our analysis we consider a generalized form of the naive
fermion
action consisting of terms which respect the hypercube structure of the
staggered fermion formulation. In addition to the standard 1-link term this
action also includes all possible 3-link terms resulting either from
introduction of a higher difference scheme for the discretization of the
fermion action or from the smearing of the 1-link term (fat-links):
\begin{eqnarray}
  \lefteqn{S_F=\sum_x~\pb(x)~\sum_\mu\gamma_\mu}\nn\\
  &&\Bigg\{
  c_{1,0}\left[U^{\rm fat}_\mu(x)~\psi(x+\muh)
    -{U_\mu^{\rm fat}}^\dagger(x-\muh)~\psi(x-\muh)\right]\label{fermact}\\
  &&  +c_{3,0}\left[U^{(3,0)}_{\mu}(x)~\psi(x+3\muh)
    -{U^{(3,0)}_{\mu}}^\dagger(x-3\muh)~\psi(x-3\muh)\right]\nn\\
  &&  +c_{1,2}\sum_{\nu\not= \mu}\left[
    U^{(1,2)}_{\mu,\nu}(x)~\psi(x+\muh+2\nuh)
    - {U^{(1,2)}_{\mu,\nu}}^\dagger(x-\muh-2\nuh)~\psi(x-\muh-2\nuh)\right. \nn \\
    &&~~~\qquad+ \left. U^{(1,-2)}_{\mu,\nu}(x)~\psi(x+\muh-2\nuh)
    - {U^{(1,-2)}_{\mu,\nu}}^\dagger(x-\muh+2\nuh)~\psi(x-\muh-2\nuh)\right]\Bigg\}\nn\\
  &&+m\sum_x\pb(x)\psi(x)~~~~,\nn
\end{eqnarray}
where $\gamma_\mu$ are the Dirac matrices and
\begin{eqnarray}
  U^{(3,0)}_{\mu}(x)&=& U_\mu(x)U_\mu(x+\muh)U_\mu(x+2\muh)~,\nn\\
  U^{(1,2)}_{\mu,\nu}(x)&=& {1 \over 2} \left[U_\mu(x)
    U_\nu(x+\muh)U_\nu(x+\muh+\nuh)
    + U_\nu(x)U_\nu(x+\nuh)U_\mu(x+2\nuh)\right]~,\nn\\
  U^{(1,-2)}_{\mu,\nu}(x) &=& {1 \over 2} \Big[U_\mu(x)U^\dagger_\nu(x+\muh-\nuh)
  U^\dagger_\nu(x+\muh-2\nuh)\nn\\
  &&\quad+~ U^\dagger_\nu(x-\nuh)
  U^\dagger_\nu(x-2\nuh)U_\mu(x-2\nuh)\Big]~,\label{fermact2}\\
  U^{\rm fat}_\mu(x)&=&{1 \over 1 + 6 \omega}
  \Bigg\{
  U_\mu(x) + \omega \sum_{\nu\not= \mu}
  \left[
    U_\nu(x)U_\mu(x+\nuh)U^\dagger_\nu(x+\muh)\right. \nn \\
  &&\quad\quad\quad\quad\quad\quad\left.
    +  U^\dagger_\nu(x-\nuh)U_\mu(x-\nuh)U_\nu(x+\muh-\nuh)
  \right]
  \Bigg\}~.\nn
\end{eqnarray}
The fat-link improved one-link term, the linear 3-link term and the angular
3-link term appear with coefficients whose dependence on the gauge coupling we
have parameterized as
follows \footnote{Note that the $N_c$ dependence of the 1-loop coefficients has been factored
out here explicitely.}
$$c_{i,j}=c^{(0)}_{i,j}+g^2~{N_c^2-1\over 2N_c}~c^{(2)}_{i,j}~.$$
Together with
the fat-link-weight $\omega$ there is thus a total number of 7 parameters in the action.
In order to reproduce the correct naive continuum limit the coefficients have to
satisfy two constraints
\begin{eqnarray}
  c^{(0)}_{1,0}+3~c^{(0)}_{3,0}+6~c^{(0)}_{1,2}&=&1/ 2~,\\
  c^{(2)}_{1,0}+3~c^{(2)}_{3,0}+6~c^{(2)}_{1,2}&=&0~.\label{1-loop_constr}
\end{eqnarray}
Our aim is to fix these coefficients such that the rotational symmetry
of the fermion propagator is improved up to one loop order.\\
 In a first step we
consider the free fermion propagator. Here we derive further constraints for
the tree level coefficients $c^{(0)}_{i,j}$.
In particular we will consider the two cases where only one or the other of the
three link terms contributes. Then the constraints fix the tree level
coefficients to certain values.\\
In the second step we calculate the self energy contributions to the fermion
propagator. Here the one-loop coefficients $c^{(2)}_{i,j}$ come into play and
can be tuned to improve also the rotational symmetry in one-loop order.\\
As gluon part of the action we choose the tree level improved 1$\times$2-action:
\begin{eqnarray}
S_G&=&{2N_c\over g^2}~\Bigg[\sum_{x, \nu > \mu}~ {5 \over 3}~\left(
1-\frac{1}{N}\re\tr\plaq_{\mu\nu}(x)\right)\nn\\
&&~\quad+~{1 \over 6}~\left(1-\frac{1}{2N}\re\tr
\left(\loOp_{\mu\nu}(x)+\lOop_{\mu\nu}(x)\right)\right)\Bigg]~~.
\end{eqnarray}\\
We note that also here the tree-level coefficients $5/3$ and $1/6$ could be
further improved at $\mathcal{O}(g^2)$. These corrections, however, are higher
order corrections for the analysis of the fermion contributions to
thermodynamic observables which we will present in the following.
\subsection{Improvement of rotational symmetry at tree level}
\label{sec_rot_inv_tree}
\begin{figure}[h]
  \centerline{
    \epsfig{file=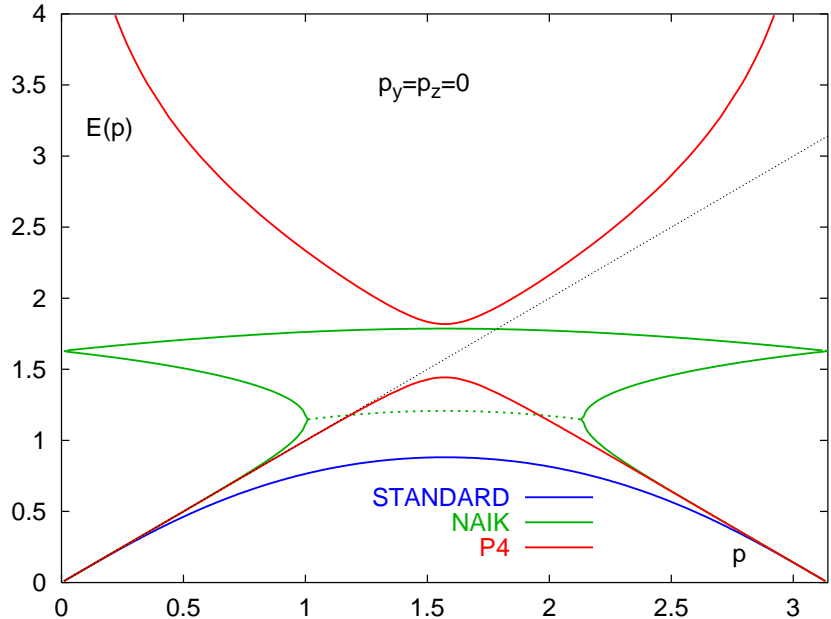,width=.8\linewidth}
    }
  \caption{
    Dispersion relation $E=E(\vec{p})$ with $\vec{p}=(p,0,0)$ for the standard staggered fermion action
    (blue), the Naik-action (green) and the p4-action (red)
    compared to the continuum dispersion relation $E=p$. Solid and dashed lines
    denote  real and complex poles of the propagator respectively.
    }
  \label{fig01}
\end{figure}
\begin{figure}[h]
  \centerline{
    \epsfig{file=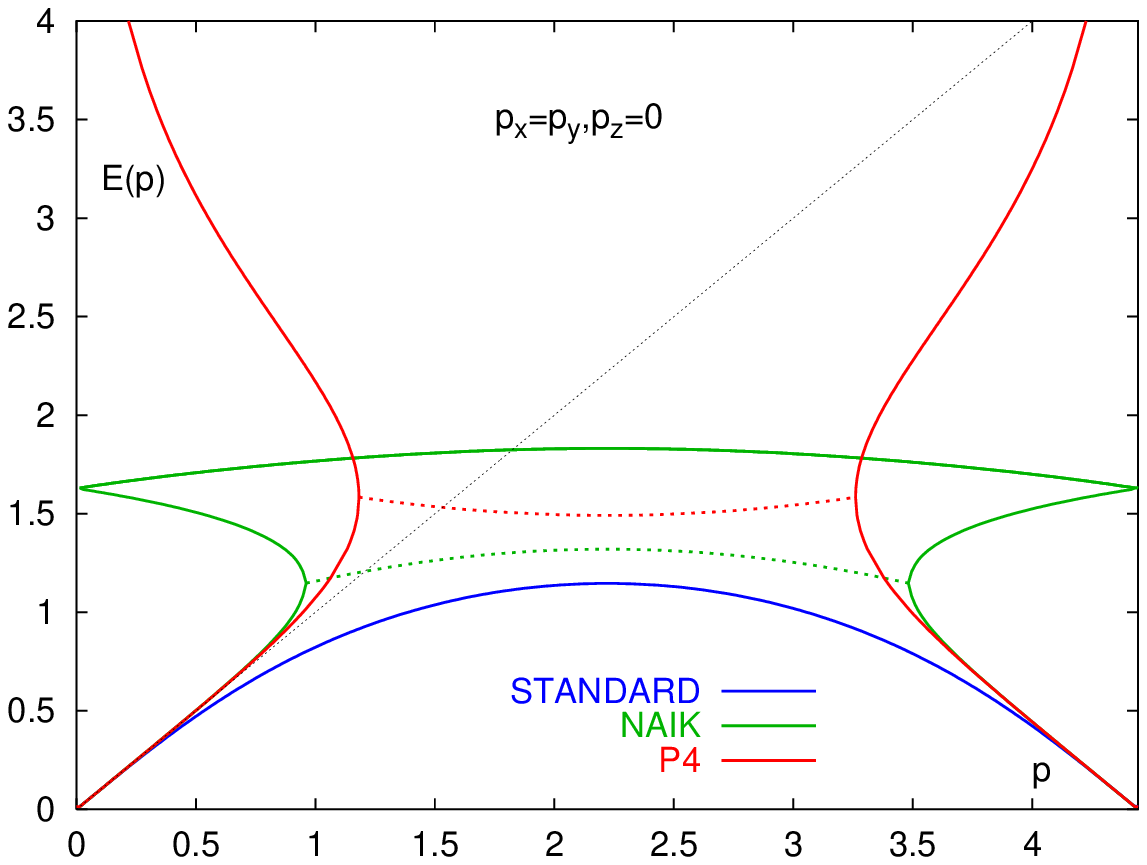,width=.8\linewidth}
    }
  \caption{
    Dispersion relation $E=E(\vec{p})$ with $\vec{p}=(p/\sqrt{2},p/\sqrt{2},0)$ for the
    standard staggered fermion action
    (green), the Naik-action (blue) and the p4-action (red)
    compared to the continuum dispersion relation $E=p$. Solid and dashed lines
    denote  real and complex poles of the propagator respectively.
    }
  \label{fig02}
\end{figure}
The inverse free fermion propagator of the action \ref{fermact} in momentum space
takes the form
\begin{equation}
{\Delta_F^{(0)}}^{-1}(p)={-i\sum_\mu\gamma_\mu h_\mu(p)+m\over
 \sum_\rho h_\rho^2(p)+m^2}
={-i\sum_\mu\gamma_\mu h_\mu(p)+m\over
D^{(0)}(p) +m^2}
~~,
\end{equation}
where the functions $h_\mu(p)$ are given in appendix \ref{app_1} (Eq. \ref{h_def}). \\
In order to obtain a rotational invariant free propagator we should achieve
that $D^{(0)}(p)=\sum_\mu~h_\mu(p)h_\mu(p)$ is a function of ${\boldmath p}^2$
only. For the standard one-link propagator of the staggered fermion action this
requirement for a rotational invariant propagator is violated at $\mathcal{O}(p^4)$.
 Expanding $D^{(0)}(p)$, \ie the trigonometric functions appearing in $h_\mu(p)$
 (see appendix \ref{app_1}), in orders of $p$ a further constraint to the
coefficients $c^{(0)}_{i,j}$ can be determined such that the free propagator is
rotational invariant up to order $p^4$, for details see appendix \ref{app_1}. The
constraint is
\begin{eqnarray}
  c^{(0)}_{1,0}+27~c^{(0)}_{3,0}+6~c^{(0)}_{1,2}=24~c^{(0)}_{1,2}~~.
\end{eqnarray}
Two simple choices which eliminate the bended or straight three link terms in
\ref{fermact}, respectively, are to set either $c^{(0)}_{1,2}\equiv 0$ which yields the
familiar {\bf Naik-action}
\begin{eqnarray}
c^{(0)}_{1,0} = {9 \over 16} ~~,~~c^{(0)}_{3,0}=-{1 \over 48}~~,
\end{eqnarray}
or to set $c^{(0)}_{3,0}\equiv 0$  which leads to
\begin{eqnarray}
c^{(0)}_{1,0} = {3 \over 8} ~~,~~c^{(0)}_{1,2}={1 \over 48}~~,
\end{eqnarray}
which we will call the {\bf p4-action}.\\
Since for the Naik-action the $\mathcal{O}(p^4)$ terms are completely
eliminated one gets even an $\mathcal{O}(a^2)$ improvement, which seems quite
accidental in this approach.

A first impression of how the improvement works is obtained by inspecting
the dispersion relation $E=E(p_x,p_y,p_z)$ resulting from the poles of the free
propagator for massless fermions ${\Delta_F^{(0)}}^{-1}$, which are solutions
of $D^{(0)}(iE,\vec{p})=0$. Figures \ref{fig01}
and \ref{fig02} show the dispersion relations for on-axis momenta
$\vec{p}=(p,0,0)$ and momenta on the planar diagonal
$\vec{p}=(p/\sqrt{2},p/\sqrt{2},0)$ respectively for the standard fermion
action, the Naik-action and the p4-action in comparison to the continuum
relation $E(p)=p$.  In case of the  Naik-action for
on-axis momenta as in case of Naik- and p4-action for planar diagonal
momenta there are also complex poles of the propagator. The real part of these
poles is plotted as thin line for distinction to the real poles.
 In the
continuum limit $a\rightarrow 0$ only the branch in the lower left corner of the $E$-$p$ plane
survives as $E=E_{\rm cont}\cdot a$ and $p=p_{\rm cont}\cdot a$. However,  the finite
cut-off $a$ leads to deviations from the continuum dispersion relation on
finite lattices. But the plots indicate that the dispersion relations for the
improved actions are close to the continuum for a much larger range of momenta
compared to the standard action. In particular the dispersion relation of the
p4-action is close to the continuum relation in nearly half the Brillouin zone for
on-axis momenta. As will be discussed in section \ref{sec3} this leads to a strong
reduction of the cut-off dependence in the ideal gas ($g\rightarrow 0$) limit.
\subsection{Improvement of rotational symmetry at ${\cal O}(g^2)$}
In order to reduce the violation of rotational symmetry also at ${\cal O}(g^2)$
we again chose to look at the fermion propagator, as it is the most fundamental
observable in the calculation of loop corrections to thermodynamic
observables. \\
Expanding the link variables $U_\mu(x)=\exp(igA_\mu(x))$ one finds an expansion
of the action in powers of g,
\begin{eqnarray}
S = S^{(0)}+gS^{(1)}+g^2S^{(2)}+g^2\tilde{S}^{(0)}+\mathcal{O}(g^3)~~.
\end{eqnarray}
Now expanding $e^{-S}$ one gets the following contributions to the fermion
  propagator in momentum space up to ${\cal  O}(g^2)$
\footnote{
Contributions arising from the expansion coefficients of the gluon action
  $S_G^{(2)}$ and ${(S_G^{(1)})}^2$ vanish because they lead to disconnected
  diagrams.
}
:
\begin{eqnarray}
  \delta_{p,q}\cdot{{\Delta_F^{(2)}}^{-1}(p)}^{ab}_{\alpha\beta}&=&\left\langle\pb_\alpha^a(q)\psi_\beta^b(p)\right\rangle\nn\\
  &=&
   \left\langle
    \pb_\alpha^a(q)\psi_\beta^b(p)
  \right\rangle_0+g^2
  \Bigg[{1\over 2}
  \left\langle
    \pb_\alpha^a(q)\psi_\beta^b(p)
    \left(
      S_F^{(1)}
    \right)^2
  \right\rangle_{0,c}\nn\\
  &&
 - \left\langle
    \pb_\alpha^a(q)\psi_\beta^b(p)S_F^{(2)}
  \right\rangle_{0,c}
 - \left\langle
    \pb_\alpha^a(q)\psi_\beta^b(p)\tilde{S}_F^{(0)}
  \right\rangle_{0,c}
  \Bigg]\nn~~,
\end{eqnarray}
where the subscripts indicate that  the expectation values are taken with respect to the
free action and that only connected parts are taken into account.
\begin{eqnarray}
  \left\langle
    \mathcal{O}
  \right\rangle_0\equiv
  {1 \over Z_0}\int~ \left[ D U\right]\left[D\pb
    D\psi\right]~\mathcal{O}~e^{-S^{(0)}}&,&
  Z_0\equiv
  \int~ \left[ D U\right]\left[D\pb
    D\psi\right]e^{-S^{(0)}}~.\nn
\end{eqnarray}
$(S_F^{(1)})^2$ and $S_F^{(2)}$ are the self-energy terms arising from the
parts independent of the 1-loop coefficients $c^{(2)}_{i,j}$, whereas
$\tilde{S}_F^{(0)}$ denotes the part depending on these. We note that
$S_F^{(0)}$ and
$\tilde{S}_F^{(0)}$ are formally the same terms apart from an exchange of
tree-level and 1-loop coefficients. The explicit form of these contributions
are given in appendix \ref{app_1}.\\
An integration over the fermion and gluon fields yields, for massless fermions,
\[
{{\Delta_F^{(2)}}^{-1}(p)}^{ab}_{\alpha\beta}  =
  \left[
    {\Delta_F^{(0)}}^{-1}(p)
    \left[
      \sum_\mu~i\gamma_\mu
      \left(
        D^{(0)}_\mu(p)+g^2~\left(\Sigma_\mu(p)+D^{(2)}_\mu(p)\right)
      \right)
    \right]
    {\Delta_F^{(0)}}^{-1}(p)
  \right]_{\alpha\beta}^{ab}~~,
\]
with
\begin{eqnarray}
D^{(0)}_\mu(p)&=&\propa~~,\nn\\
D^{(2)}_\mu(p)&=&\propb~~,\nn\\
\Sigma_\mu(p)&=&\sigmaa+\sigmab\nn~~.
\end{eqnarray}
The explicit results are given in appendix \ref{app_3}.
The self-energy term contains, besides the usual continuum contribution
$(S_F^{(1)})^2$, also a 1-loop term , $S_F^{(2)}$, which is a pure lattice
artifact. The $D^{(2)}$ term, which arises from the non-zero 1-loop
coefficients $c^{(2)}_{i,j}$, acts
like a
counter-term in the sense that it ought to adjust the 1-loop contributions,
although it does not remove any divergence.\\
To achieve a rotational invariant fermion propagator to $\mathcal{O}(g^2)$ we
consider
the denominator of ${\Delta_F^{(2)}}^{-1}(p)$, which can be written as
\begin{eqnarray}
  \mathcal{D}(p)&\equiv&\sum_\mu\left(D^{(0)}_\mu(p)+g^2~
\left(\Sigma_\mu(p)+D^{(2)}_\mu(p)\right)\right)^2\nn\\
  &=&\sum_\mu D^{(0)}_\mu(p)D^{(0)}_\mu(p) + 2 g^2\sum_\mu
D^{(0)}_\mu(p)\left(\Sigma_\mu(p)+D^{(2)}_\mu(p)\right)+\mathcal{O}(g^4)~~,\nn
\end{eqnarray}
and demand that $\mathcal{D}$ should be a function of $p^2$ only,
analogous to the tree level case discussed in section \ref{sec_rot_inv_tree}.
By this demand one obtains constraints for the 1-loop
coefficients $c^{(2)}_{i,j}$ analogous to the approach in the previous
section. But in contrast to the tree level case a straight forward expansion in
powers of $p$ is not possible due to the logarithmic divergence in the
contributions arising from
$(S_F^{(1)})^2$.\\
Alternatively we consider on-axis and off-axis momenta having the same
magnitude, $p_1=(\tilde{p},0,0,0)$ and
$p_2=(\tilde{p}/\sqrt{2},\tilde{p}/\sqrt{2},0,0)$, and then determine the
coefficients by solving the equation
\footnote{
The term $\sum_\mu D^{(0)}_\mu(p)D^{(0)}_\mu(p)$ is already rotational
symmetric to order $p^4$ with the tree-level coefficients derived in
section \ref{sec_rot_inv_tree}. Therefore only the term proportional to $g^2$
has to be taken into account.
}
\begin{equation}
\mathcal{D}(p_1)=\mathcal{D}(p_2)\label{d_eq_d}
\end{equation}
 for small momenta $\tilde{p}$.\\
In general $\mathcal{D}(p)$ depends also on the fat-link-weight $\omega$.
But since the fat-link-weight has a negligible effect on thermodynamics,
as it will turn out in the next section, where we look at the free
energy density at $\mathcal{O}(g^2)$, we fix the value of the fat-link-weight
to $\omega$=0 for this calculation.
The restriction to one or the other 3-link term in the
action together with the constraint \ref{1-loop_constr} leaves only one free 1-loop parameter,
which can be determined as solution of equation \ref{d_eq_d}.

\begin{figure}[t]
  \epsfig{file=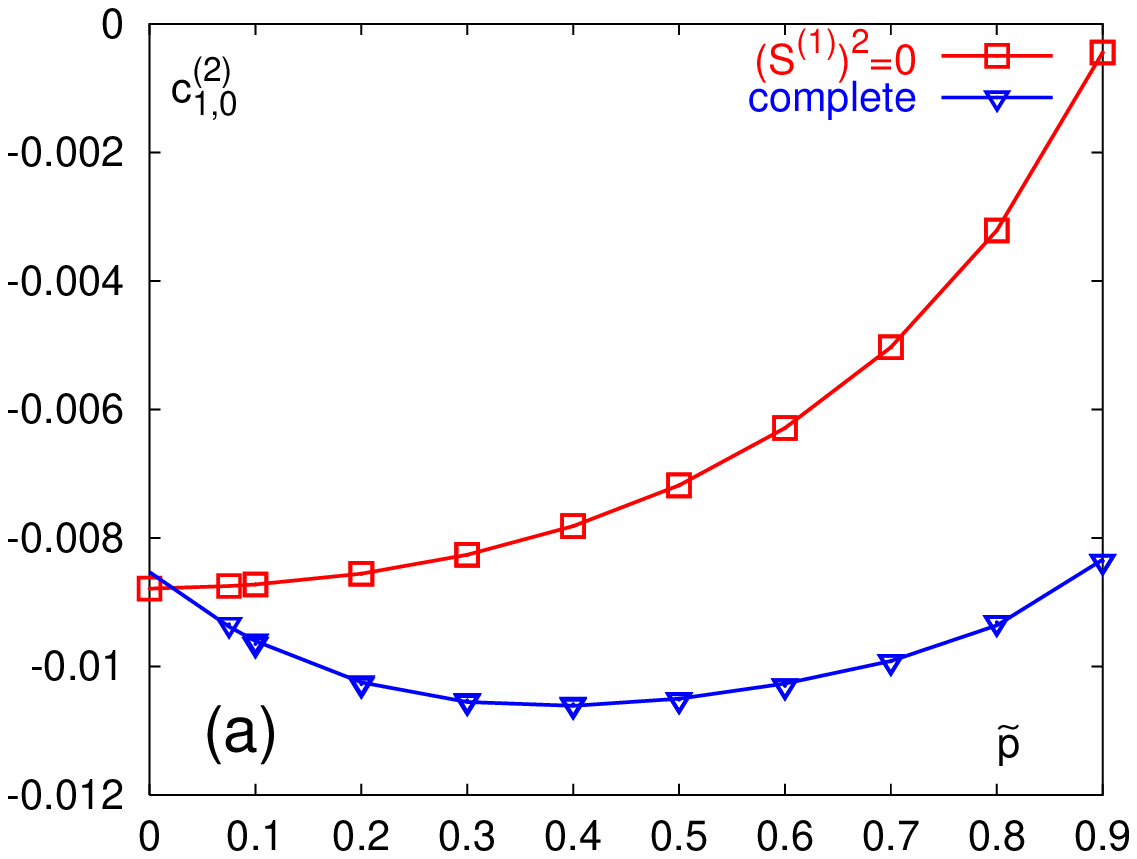,width=.5\linewidth}
  \epsfig{file=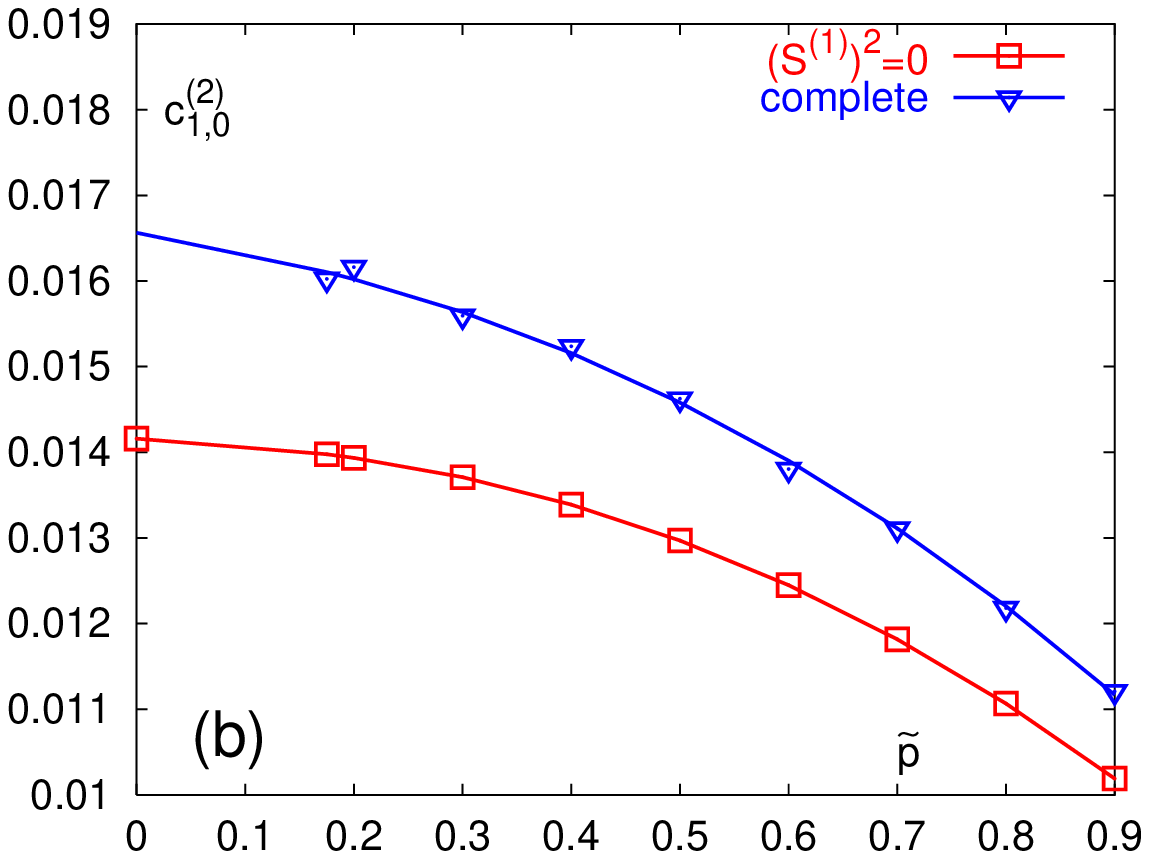,width=.5\linewidth}
  \caption{
    Solution of the equation \ref{d_eq_d} with
    complete contributions (blue) and without $(S_F^{(1)})^2$ contribution
    (red) for (a) the Naik-action and (b) the p4-action. Lines in case
    of the complete contributions are extrapolated by a polynomial fit.
    }
  \label{fig1}
\end{figure}
 The
$\tilde{S}^{(0)}$ term is a simple function of $p$.
In the different parts of the $S_F^{(2)}$ contribution the $p$ dependence can
be factored out of the gluonic integrals. Thus, they can be computed once
for all values of $p$.
In appendix \ref{app_3} we show that only three distinct
gluonic integrals appear in the $S_F^{(2)}$ contributions.
These gluon-integrals have been calculated to high accuracy using
Gauss-Legendre integration.\\
In principle one can expand these contributions in
powers of $p$ as in the previous section, but in contrast to that the
$(S_F^{(1)})^2$ contribution can not be expanded due to a logarithmic
divergence in the $p\rightarrow 0$ limit. The gluon-loops have to be calculated
for each of the momenta $\tilde{p}$ separately, since there is no factorization
in gluonic and fermionic parts. The mentioned divergence causes also numerical
difficulties for small fermion momenta $p$, where the pole of the inner gluon
line, which is at $p-k=0$, runs into the pole of the gluon propagator at
$k=0$. Therefore some care has to be taken to get reliable results. In practice
this problem prohibits to perform our numerical integrations at arbitrary small
$\tilde{p}$.

Solutions of the equation \ref{d_eq_d} are plotted in Figure \ref{fig1} for the
Naik-action and the p4-action as function of the momentum $\tilde{p}$. The
squares are solutions for $c_{1,1}^{(2)}$, one obtains including only the
$D^{(2)}$ term and the $S_F^{(2)}$ part of $\Sigma$, \ie  setting
$(S_F^{(1)})^2$ to zero. Since the  $p$ dependence of these parts is known
explicitely we get precise results down to very small momenta.
The triangles represent the solutions including all
contributions. The scattering of points at small $\tilde{p}$ is due to the
numerical difficulties in this regime for the $(S_F^{(1)})^2$ part. In the Naik
case we obtain stable results for $\tilde{p}\ge 0.1$, whereas in the p4 case
the results are trustable only for $\tilde{p}\ge 0.3$. In these regimes
we have increased the number of Gaussian points until the values became stable,
\ie up to 320 Gaussian points in each direction for the small momenta. \\
These
plots indicate that the main contribution to the violation of rotational
symmetry originates from the $S_F^{(2)}$ part of the self-energy term which is
quite reasonable since this term is a pure lattice artifact.

\noindent Using a polynomial fit we get the extrapolated values of the
coefficients:

\begin{tabular}{rclcD{.}{.}{5}clcD{.}{.}{5}}
{\rm Naik-action}&:&$c^{(2)}_{1,0}$&=&-0.0085&,&$c^{(2)}_{3,0}$&=&0.00283\\
{\rm p4-action}&:&$c^{(2)}_{1,0}$&=&0.0165&,&$c^{(2)}_{1,2}$&=&-0.00275\\
\end{tabular}

\noindent It is remarkable that these coefficients indicate an over-improvement using the
tree-level improvement at finite $g$, as the tree-level coefficients are
corrected towards the standard staggered action.

\section{Free energy in 1-loop perturbation theory}
\label{sec3}

Thermodynamic observables show large cut-off dependencies in finite temperature
lattice simulations.  A suitable quantity to study the influence of finite
cut-off effects on thermodynamic observables is the perturbative high
temperature limit which is known from continuum perturbation theory up to order
$g^2$ for a long time \cite{Kap79}\footnote{
All perturbatively calculable terms (up to $\mathcal{O}(g^5)$) have been
calculated only recently for the pure gauge sector \cite{Arn94}.
}.
 The free energy density for $n_f$ flavours of massless
quarks and the $SU(N_c)$ colour group is
\begin{eqnarray}
f_{\rm cont}
&\equiv& f^{(0)}_{{\rm cont},G}+f^{(0)}_{{\rm cont},F}
+g^2\left(f^{(2)}_{{\rm cont},G}
+f^{(2)}_{{\rm cont},F}\right)+{\cal O}(g^3)~~,{\rm where}\nn\\
f^{(0)}_{{\rm cont},G}/T^4&=&-{1\over 45}\pi^2\left( N_c^2-1 \right)~~,\nn\\
f^{(2)}_{{\rm cont},G}/T^4&=&{1 \over 144} g^2 \left( N_c^2-1 \right)N_c~~,\nn\\
f^{(0)}_{{\rm cont},F}/T^4&=&- n_f {7 \over 180}\pi^2N_c~~,\nn\\
f^{(2)}_{{\rm cont},F}/T^4&=&n_f{5 \over 576} \left( N_c^2-1 \right)~~.
\label{cont_f}
\end{eqnarray}
Up to this order the
thermodynamic relation $\epsilon=3p=-3f$ holds, where $\epsilon$ is the
energy density, $p$ is the pressure and $f$ is the free energy density.\\
On lattices with temporal extension $N_\tau$ large ${\cal O}(1/N_\tau^2)$, \ie
$\mathcal{O}((aT)^2)$,
 deviations from \ref{cont_f}
are found in the gluonic contribution for the standard plaquette action
as well as in the fermionic part for the standard Kogut-Susskind action.\\
Tree level improved gauge actions reduce
these cut-off effects in the leading order perturbative ideal gas limit.
It has been shown that the improvement persists at finite temperatures
well below those at which the ideal gas limit is approached \cite{Bei96}.
In the fermionic sector thermodynamic observables  show even
larger deviations from the continuum Stefan-Boltzmann limit using the standard
staggered fermion formulation \cite{Eng96}. By construction the Naik-action also
reduces these deviations in the fermionic contributions to the energy density
$\epsilon$ in leading order. \\
At 1-loop order it is known that the standard staggered action
also leads to large cut-off dependent deviations in the energy density
$\epsilon$ \cite{Hel85}. We will analyze in the following in how far these are
reduced for the improved actions constructed in the previous section.

We calculate the fermionic contribution to the free energy
density $f_F$ in lattice perturbation theory up to order $g^2$ for various fermion
actions introduced in the previous section including the 1-loop improved
actions and fat-link improvement. Only at this order will the influence of
fat-link improvement, 1-loop coefficients and also the chosen gauge-action show
up.\\
 Since we are dealing with naive
fermions the
number of flavours is $n_f=16$.
Up to 1-loop order lattice perturbation theory is equivalent for naive and
staggered fermions and the correct $n_f$ dependence which only is a
multiplicative factor at this order, can easily be introduced at the end
\cite{Sha81}.
 We always chose the bare mass $m=0$ to get
comparable results.\\
The reason that we chose the free energy density $f$ to look at is that it is
defined by a very simple relation, just the logarithm of the partition
function, instead of a derivative of the logarithm of the partition in case of
the energy density $\epsilon$:\\
\begin{eqnarray}
  e^{-{V\over T}f_F}&\equiv& Z = \int~ \left[ D U\right]\left[D\pb
    D\psi\right]e^{-S_F}\nn\\
  &=& Z_0 \Bigg\{1
    -g^2\left[
      \left\langle
        S_F^{(2)}
      \right\rangle_0
      -{1\over 2}\left\langle
        \left(S_F^{(1)}\right)^2
      \right\rangle_0
    +  \left\langle
        \tilde{S}_F^{(0)}
      \right\rangle_0
    \right]+\mathcal{O}(g^3)
  \Bigg\}~~.\label{def_f}
\end{eqnarray}
Here again we expanded the fermion action in powers of $g$. The subscript zero
is defined as in the previous section and pure gluonic contributions are
neglected also. Taking the logarithm of \ref{def_f} one finds:
\begin{eqnarray}
f_F&=&f^{(0)}_F+g^2f^{(2)}_F+\mathcal{O}(g^3)\nn\\
&=&f^{(0)}_F+g^2{T \over V}
\left[
      \left\langle
        S_F^{(2)}
      \right\rangle_0
      -{1\over 2}\left\langle
        \left(S_F^{(1)}\right)^2
      \right\rangle_0
    +  \left\langle
        \tilde{S}_F^{(0)}
      \right\rangle_0
    \right]+\mathcal{O}(g^3)~.
\end{eqnarray}
Explicit expressions for the expansion coefficients are given in
appendix \ref{app_4}. The various contributions correspond to the following diagrams:\\
\parbox{.5\linewidth}{
\begin{eqnarray}
f^{(0)}_F&=&\fa~~~,\nn\\
\left\langle\left( S_F^{(1)}\right)^2\right\rangle_0&=&\ssb~,\nn
\end{eqnarray}
}
\parbox{.5\linewidth}{
\begin{eqnarray}
\left\langle\tilde{S}_F^{(0)}\right\rangle_0&=&\ssa~,\nn\\
\left\langle S_F^{(2)}\right\rangle_0&=&\ssc~.\nn
\end{eqnarray}
}\\
\begin{figure}[t]
  \centerline{
    \epsfig{file=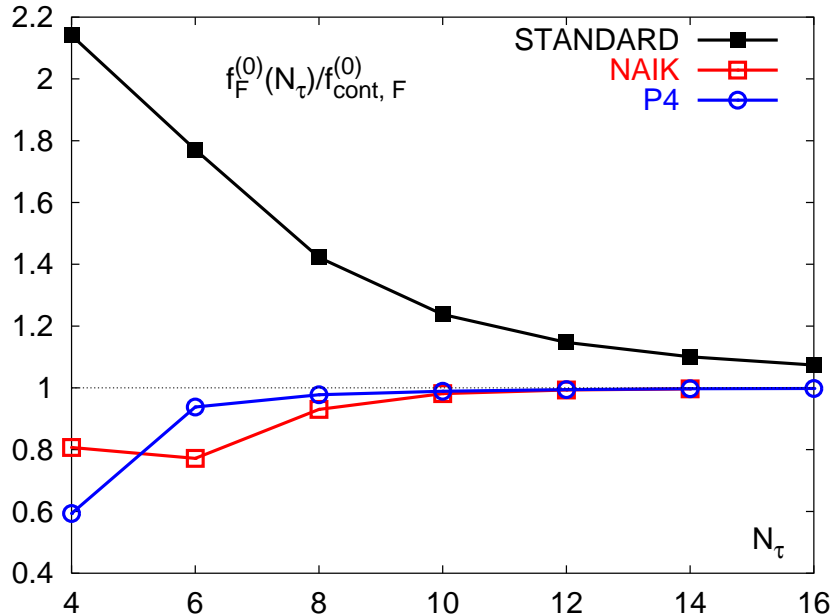,width=.8\linewidth}
    }
  \caption{
    Fermionic tree level contribution to the free energy density normalized to
    the fermionic continuum tree level contribution as function of temporal
    extension
    $N_\tau$ for standard Kogut-Susskind action (black), Naik-action
    (red) and p4-action (blue).
    }
  \label{fig2}
\end{figure}
\begin{figure}[t]
  \centerline{
    \epsfig{file=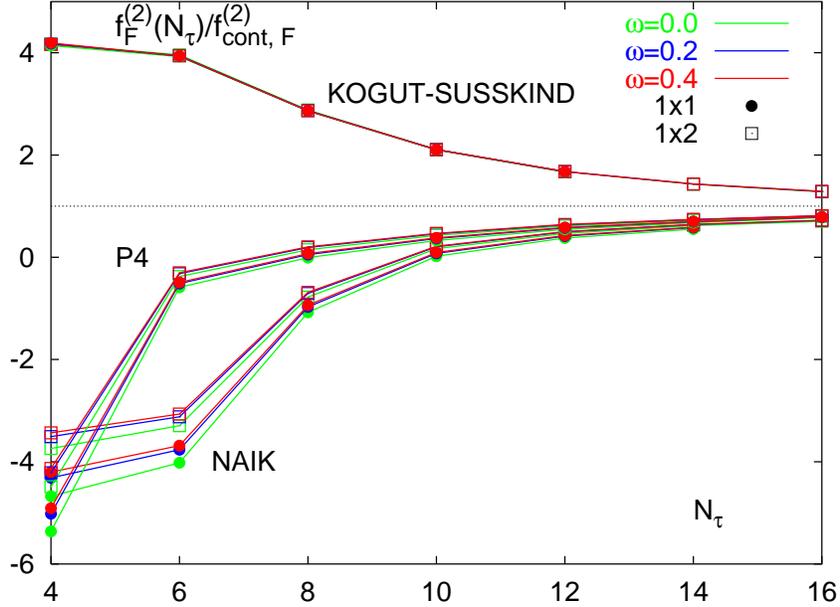,width=.8\linewidth}
    }
  \caption{
    Fermionic 1-loop contribution to the free energy density normalized to the
    fermionic continuum 1-loop  contribution as function of temporal extension
    $N_\tau$ for various fermion actions. Different colours represent the
    value of the fat-link parameter $\omega=0.0,~0.2,~0.4$. Circles denote the
    use of the standard gluon action, open squares the use of the 1$\times$2
    gluon action. In the Naik- and p4-action only
    tree-level coefficients are included, \ie $c_{i,j}^{(2)}\equiv 0$. The
    values for $N_\tau>8$ are extrapolated.
    }
  \label{fig3}
\end{figure}
\begin{figure}[t]
  \centerline{
    \epsfig{file=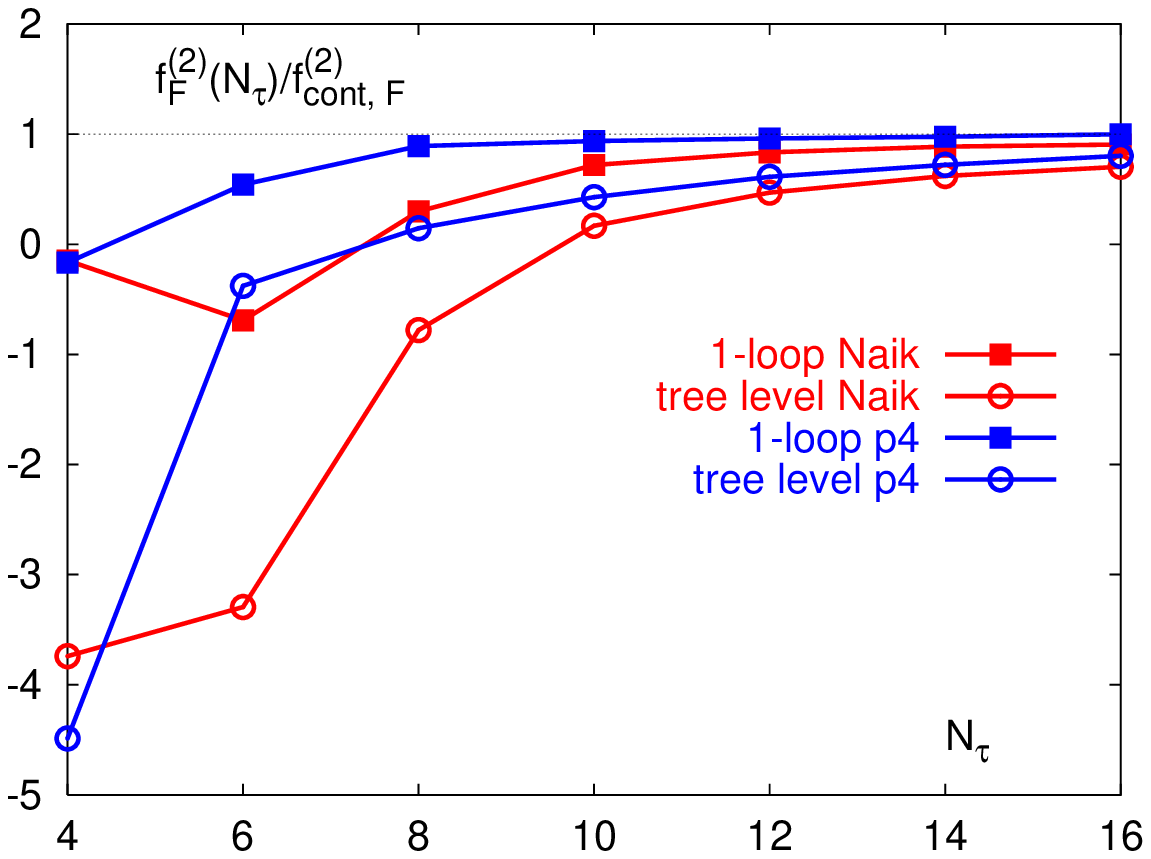,width=.8\linewidth}
    }
  \caption{
    Fermionic 1-loop  contribution to the free energy density normalized to the
    fermionic continuum tree level contribution as function of temporal
    extension $N_\tau$ for the Naik-action (red) and the p4-action
    (blue) including only tree-level improvement, \ie
    $c_{i,j}^{(2)}\equiv 0$, (open circles) and including also 1-loop
    improvement (squares). Always the 1$\times$2 gluon action is used.  The
    values for $N_\tau>8$ are extrapolated.
    }
  \label{fig4}
\end{figure}
Obviously there is a correspondence between these contributions and those to
the fermion propagator discussed in the previous section. One gets the graphs
just by connecting the incoming and outgoing fermion lines of the diagrams of
the fermion propagator. The $\tilde{S}_F^{(0)}$ contribution includes the whole
dependence on the 1-loop coefficients and does not contribute if these are
ignored. The $S_F^{(2)}$ term again can be factorized in fermionic
and gluonic loop integrals in contrast to the $( S_F^{(1)})^2$ term.\\
In order to extract the finite temperature contribution we calculated the
difference of the free energy density on $N_\sigma^3\times N_\tau$ and
$N_\sigma^4$ lattices. The spatial extension was set to infinity,
$N_\sigma=\infty$. The loop integrals over spatial momenta were done
using Gauss-Legendre integration. For the zero temperature contributions
also the temporal momenta were integrated over with Gauss-Legendre
integration, while for the finite temperature contributions the finite
sums over temporal momenta were carried out explicitly.\\
The  numerical effort to calculate  the 2-loop integrals in particular for the
$(S_F^{(1)})^2$ term is large and grows like $(\#$Gauss-points$)^{8}$, therefore the number of Gaussian points was
at maximum 32 in each direction, which is a total number of about $10^{12}$
points. For small temporal extension $N_\tau\le 8$ one finds a good
convergence of the 2-loop contributions increasing the number of Gaussian
points up to 32. For larger values of $N_\tau$, however, the limit is not
reached with this maximum number of points. However, since the results show a clear
$(\#$Gauss-points$)^{-4}$ behaviour they can be extrapolated to infinite number
of Gaussian points also for $N_\tau> 8$. Of course the uncertainty of the
extrapolation grows with increasing $N_\tau$, but since our main interest lies
in the deviations from continuum perturbation theory for small values of
$N_\tau$ this does not affect our conclusions.

Figure \ref{fig2} shows the deviations of the fermionic tree-level
contributions to the free energy density from the Stefan-Boltzmann limit for
the standard staggered fermion action, the Naik-action and the p4-action. The
improved actions reduce the large cut-off dependent deviation of the standard
action drastically from more than 20 at $N_\tau$=4 to about 20\% in case of
the Naik-action and about 40\% in case of the p4-action. At $N_\tau$=6 and
greater the improvement works even better. Here the deviations for the
p4-action are below a few percent compared to nearly 80\% for the standard
action at $N_\tau$=6.\\
The results for the fermionic part of the free energy density at order $g^2$, $f^{(2)}$,
normalized to the continuum order $g^2$ contribution, $f_{\rm cont}^{(2)}$, are
plotted in Figures \ref{fig3} and \ref{fig4}. Figure \ref{fig3} shows the
influence of fat-link improvement and the chosen gauge action for the
considered tree-level improved and the standard staggered fermion actions, \ie
1-loop coefficients are set to zero. For all actions large deviations from
the continuum value are found at small values of $N_\tau$, although the
p4-actions approach the continuum value most rapidly. The main differences are
caused by the
choice of tree-level coefficients of the fermion action rather than the gauge
action or the fat-link parameter $\omega$.\footnote{The values of $\omega$ are chosen in the same range as used
by the MILC collaboration for investigation of flavour symmetry breaking
\cite{fat}.}
This is not unexpected, because fat-link improvement affects the infrared
behaviour and reduces flavour symmetry
breaking rather than the ultraviolet regime which is responsible for the  high
temperature behaviour. Further, improved gauge actions are designed to reduce
cut-off distortions in the gluonic contributions, but they have shown to reduce
flavour symmetry breaking also \cite{Ber97}, which indicates rather
an influence on infrared sensitive fermionic observables.

The influence of 1-loop improvement of Naik- and p4-action provided in the
previous section on the high temperature behaviour at order $g^2$ is shown in
Figure \ref{fig4}. Here in all cases the 1$\times$2 gauge action is used and the
fat link parameter equals zero. The plot indicates a large reduction of cut-off
effects at order $g^2$ for both types of 1-loop improved fermion actions compared
to the corresponding  tree-level action. Already at $N_\tau$=6 the deviations
are reduced to less than 50\% for the 1-loop p4-action and to about 140\% for
the 1-loop Naik-action, where the 1-loop contribution for the standard
staggered action is about 4 times larger than the continuum 1-loop
value.

As stated in the previous section, the 1-loop coefficients correct the
tree-level coefficients towards the standard action in both cases, Naik-
and p4-action. In correspondence to that we observe a change of the sign of
$f_F^{(2)}$ comparing standard and tree-level improved actions, \ie an
``over-improvement'' concerning the order $g^2$, which is reduced when including the
1-loop coefficients.

\section{Conclusion}
In this work we have presented a perturbative construction of staggered
fermion actions with improved rotational symmetry of the fermion
propagator at tree-level as well as in 1-loop order. In particular we considered
tree-level and 1-loop improved versions of the Naik- and the p4-action
both with and without additional fat-link improvement.\\
The tree-level improvement shows up in the dispersion relation which is close
to the continuum relation in a much larger momentum
range for the improved actions
than for the standard action.\\
A perturbative calculation of the fermionic tree-level and 1-loop contributions
to the free energy density $f$ has been performed to investigate the influence
of improved rotational symmetry and fat-link improvement on the high
temperature behaviour. At tree level as well
as in 1-loop order the standard staggered
action leads to large cut-off dependent deviations from the perturbative
continuum value. We find a large reduction of these ultraviolet cut-off effects
at tree-level for the tree-level improved actions. For example at $N_\tau=4$
the remaining cut-off effects are about a factor 10 smaller than the
cut-off effects of the standard action.\\
Of course the coefficients of the $\mathcal{O}(a^2)$ deviations  depend on the
observable.
An earlier calculation of fermionic contributions to the energy density
$\epsilon$  at tree-level has shown a reduction of cut-off distortions for the
p4- and the Naik-action which is comparable to the results for the free energy
density we present in this work although in case of the energy density one
finds even smaller deviations for the p4-action of about 1\% at $N_\tau$=4
\cite{Bei98}.

Tree-level improvement,
however, does not help at 1-loop order as expected. Our analysis showed that
cut-off distortions are of comparable magnitude for the standard action as for
the tree-level improved actions in the 1-loop contributions to the free energy
density. It also turned out that fat-link improvement has a rather negligible
influence on the high temperature behaviour, \ie it neither reduces nor
increases the cut-off effects significantly.
This is not surprising since fat-link improvement was designed to affect
the infrared rather than the ultraviolet behaviour.\\
Finally we find a large reduction of these ultraviolet cut-off
effects at 1-loop order for the actions with improved rotational symmetry up to
$\mathcal{O}(g^2)$. Already at $N_\tau$=6 the deviations
are reduced to less than half of the continuum 1-loop
value for the 1-loop p4-action and to about $1.4$-times the continuum value for
the 1-loop Naik-action, where the 1-loop contribution for the standard
staggered action is about 4 times larger than the continuum 1-loop
value.\\
Therefore the 1-loop improved p4-action in combination with
fat-links is a particularly suitable candidate for finite temperature simulations
in full QCD, combining the advantages of an improved high temperature behaviour
up to 1-loop order with an improved flavour symmetry. And last but not least
it retains the simplicity of the contributing local operators.

\noindent{\Large \bf Acknowledgments} ~~This work was partly supported by the TMR network
{\em Finite Temperature Phase Transitions in Particle Physics}, EU contract
no. ERBFMRX-CT97-0122. The work of U.M.H. was partly supported by DOE contracts
DE-FG05-85ER250000 and DE-FG05-96ER40979, and he
gratefully acknowledges support from the Zentrum
f\"ur Interdisziplin\"are Forschung, Universit\"at Bielefeld.

\newpage

\begin{appendix}
\section{Appendix}
We give here explicit expressions as well as some technical details of the
perturbative calculations. It is organized as follows: In part \ref{app_1} we
define the fundamental terms and functions we refer to in the following
parts. Part \ref{app_2} and part \ref{app_3} contain some details of the
determination of the tree-level and 1-loop coefficients respectively. In part
\ref{app_4} we give the explicit results of the 1-loop calculation of the
fermionic contribution to the free energy density.
\subsection{General 1-loop results and definitions}
\label{app_1}

For the trigonometric functions we use the short hand notation:
\begin{eqnarray}
s_\mu(p)&\equiv&\sin(p_\mu)~~,\nn\\
c_\mu(p)&\equiv&\cos(p_\mu)~~.\nn
\end{eqnarray}
The inverse gluon propagator of the 1$\times$2 action was calculated in
\cite{Bei96}. For the standard plaquette action as for the 1$\times$2 action
the propagator can be written as
\begin{eqnarray}
  {\Delta_G}_{\mu,\nu}(k)&=&
  {D_G}_\mu(k)\delta_{\mu,\nu}-{E_G}_{\mu,\nu}(k)+\xi~g_\mu(k)g_\nu(k)~,\nn\\
  {D_G}_\mu(k)&=&4~a_{1,1}~\sum_\nu s_\nu^2(k/2)-16~a_{1,2}~\sum_\nu s_\nu^2(k/2)
  \left(2-s_\nu^2(k/2)-s_\mu^2(k/2)\right)~,\nn\\
  {E_G}_{\mu,\nu}(k)&=&4~a_{1,1}~s_\mu(k/2)s_\nu(k/2)\nn\\
  &&-16~a_{1,2}~ s_\mu(k/2)s_\nu(k/2)
  \left(2-s_\nu^2(k/2)-s_\mu^2(k/2)\right)~~,\nn
\end{eqnarray}
with $a_{1,1}\equiv 1$, $a_{1,2}\equiv 0$ for the standard Wilson one-plaquette action and
$a_{1,1}\equiv 5/3$, $a_{1,2}\equiv -1/12$ for the 1$\times$2 action.
In both cases we choose the gauge fixing term $g_\mu(k)=2s_\mu(k/2)$ and
Feynman gauge $\xi=1$.

The basis of our 1-loop calculations is an expansion of the fermion action in
powers of the bare coupling $g$:
\begin{eqnarray}
S_F = S_F^{(0)}+gS_F^{(1)}+g^2S_F^{(2)}+g^2\tilde{S}_F^{(0)}+\mathcal{O}(g^3)~~.
\end{eqnarray}
To achieve that, we expand the exponential representation of the link variables
$U_\mu(x)=\exp(igaA_\mu(x))$, with the gauge fields  $A_\mu\equiv
\sum_{b=1}^{N_c^2-1}A_\mu^b\lambda^b$ and normalization $2\tr
\lambda^a\lambda^b=\delta_{ab}$ for the group generators $\lambda^a$. The lattice spacing is set to $a=1$.
In momentum space we find then
\begin{eqnarray}
  S_F^{(0)}&=&i\int_p\bar{\psi}(p)\left\{
      \sum_\mu\gamma_\mu h_\mu(p)+m
    \right\}\psi(p)~,\nn\\
  \tilde{S}_F^{(0)}&=&i{N_c^2-1\over 2N_c}\int_p\bar{\psi}(p)~
      \sum_\mu\gamma_\mu h^{(2)}_\mu(p)
    ~\psi(p)~,\nn\\
  S_F^{(1)}&=&i\int_p\int_k\bar{\psi}(p)~
      \sum_\mu\gamma_\mu \hat{S}^{(1)}_\mu(p,k)
    ~\psi(p-k)~,\nn\\
  S_F^{(2)}&=&-{i \over 2}\int_p\int_{k_1}\int_{k_2}\bar{\psi}(p)~
      \sum_\mu\gamma_\mu \hat{S}^{(2)}_\mu(p,k_1,k_2)
    ~\psi(p-k_1-k_2)~,\nn
\end{eqnarray}
where
\begin{eqnarray}
\hat{S}^{(1)}_\mu(p,k)&=&\sum_\rho~K_{\mu;\rho}(p,k)~A_\rho(k)~,\nn\\
\hat{S}^{(2)}_\mu(p,k_1,k_2)&=&\sum_{\rho,\sigma}~
L_{\mu;\rho,\sigma}(p,k_1,k_2)~A_\rho(k_1)A_\sigma(k_2)~~.\nn
\end{eqnarray}
Here we use the following definitions also referred to in appendices \ref{app_2} -
\ref{app_4}:
\begin{eqnarray}
  h_\mu(p)&=&2~s_\mu(p)\Big[
    c_{1,0}^{(0)}+2 ~c_{1,2}^{(0)}\sum_{\nu \not= \mu}c_\nu(2p)
  \Big]
  +2~c_{3,0}^{(0)}s_\mu(3p)~,\label{h_def}\\
  h^{(2)}_\mu(p)&=&2~s_\mu(p)\Big[
    c_{1,0}^{(2)}+2 ~c_{1,2}^{(2)}\sum_{\nu \not= \mu}c_\nu(2p)
  \Big]
  +2~c_{3,0}^{(2)}s_\mu(3p)~,\nn\\
  K_{\mu;\rho}(p,k)&=&
  c_{1,0}^{(0)}~\mathcal{A}^{\rm fat}_{\mu;\rho}(\omega;p,k)
  +c_{3,0}^{(0)}~\mathcal{A}^{(3,0)}_{\mu;\rho}(p,k)
  +c_{1,2}^{(0)}~\mathcal{A}^{(1,2)}_{\mu;\rho}(p,k)~,
  \nn\\
  L_{\mu;\rho,\sigma}(p,k_1,k_2)&=&
   c_{1,0}^{(0)}~\mathcal{B}^{\rm fat}_{\mu;\rho,\sigma}(\omega;p,k_1,k_2)
  +c_{3,0}^{(0)}~\mathcal{B}^{(3,0)}_{\mu;\rho,\sigma}         (p,k_1,k_2)
  +c_{1,2}^{(0)}~\mathcal{B}^{(1,2)}_{\mu;\rho,\sigma}         (p,k_1,k_2)~,
\nn
\end{eqnarray}
where
\begin{eqnarray}
  \mathcal{A}^{\rm fat}_{\mu;\rho}(\omega;p,k)&=&
  2~ c_\mu(p-k/2)\nn\\
  &&
    \hspace*{-1cm}
    \cdot
    \Bigg[
      \delta_{\mu,\rho}-{4\omega\over 1+6\omega}
      \Bigg(
        \delta_{\mu,\rho}\sum_{\nu\not= \mu}s_\nu^2(k/2)
        -(1-\delta_{\mu,\rho})~s_\mu(k/2)s_\rho(k/2)
      \Bigg)
    \Bigg]~,
  \nn\\
  \mathcal{B}^{\rm fat}_{\mu;\rho,\sigma}(\omega;p,k_1,k_2)&=&
  2~ s_\mu(p-k_1/2-k_2/2)\nn\\
  &&
    \hspace*{-1cm}\cdot\Bigg[
  \delta_{\mu,\rho}\delta_{\mu,\sigma} -{4\omega\over 1+6\omega}
  \Bigg(
  \delta_{\mu,\rho}\delta_{\mu,\sigma}\sum_{\nu\not= \mu}
  s_\nu^2(k_1/2+k_2/2)
  \nn\\
  &&\qquad
  +2~(1-\delta_{\mu,\rho})~\delta_{\rho,\sigma}~s_\mu(k_1/2)s_\mu(k_2/2)
  c_\rho(k_1/2+k_2/2)
  \nn\\
  &&\qquad
  - \delta_{\mu,\rho}~ (1-\delta_{\mu,\sigma})~
 \{s_\mu(k_2/2) - i c_\mu(k_2/2) \} ~s_\sigma(k_1+k_2/2)
  \nn\\
  &&\qquad
  - \delta_{\mu,\sigma}~ (1-\delta_{\mu,\rho})~
 \{s_\mu(k_1/2) + i c_\mu(k_1/2) \} ~s_\sigma(k_2+k_1/2)
  \Bigg)    \Bigg]~,
\nn\\
  \mathcal{A}^{(3,0)}_{\mu;\rho}(p,k)&=&2~\delta_{\mu,\rho}~c_\mu(3p-3k/2)\left(3-4s_\mu^2(k/2)\right)\nn
  ~,\nn\\
  \mathcal{B}^{(3,0)}_{\mu;\rho,\sigma}         (p,k_1,k_2)
  &=&2~\delta_{\mu,\rho}~\delta_{\mu,\sigma}~
  s_\mu(3p-3k_1/2-3k_2/2)\left[
    9-4s_\mu^2(k_1/2+k_2/2)\right.\nn\\
  &&\qquad\qquad\qquad\left.-4s_\mu^2(k_1/2)-4s_\mu^2(k_2/2)-4s_\mu^2(k_1/2-k_2/2)
  \right]
  \nn\\
  \mathcal{A}^{(1,2)}_{\mu;\rho}(p,k)&=&4~c_\mu(p-k/2)
    ~\delta_{\mu,\rho}\sum_{\nu\not= \mu}c_\nu(2p-k)c_\nu(k)
\nn\\
  &&-8~(1-\delta_{\mu,\rho})~s_\mu(p-k/2)c_\mu(k/2)s_\rho(2p-k)c_\rho(k/2)~,
  \nn\\
\mathcal{B}^{(1,2)}_{\mu;\rho,\sigma}         (p,k_1,k_2)
  &=&4~s_\mu(p-k_1/2-k_2/2)\nn\\
  &&
    \hspace*{-1cm}\cdot
  \Bigg[
  \delta_{\mu,\rho}~\delta_{\mu,\sigma}\sum_{\nu\not= \mu}
  c_\nu(k_1+k_2)c_\nu(2p-k_1-k_2)
  \nn\\
  && \hspace*{-.5cm}+2~ (1-\delta_{\mu,\rho})~\delta_{\rho,\sigma}~c_\mu(k_1/2+k_2/2)
  \nn\\
  &&  \hspace*{-.5cm}\cdot\Big(
    c_\rho(2p-3k_1/2-k_2/2)+c_\rho(2p-k_1-k_2)c_\rho(k_1/2+k_2/2)
  \Big)
  \Bigg]\nn\\
  &&+8~\delta_{\mu,\rho}~(1-\delta_{\mu,\sigma})
  c_\rho(p-k_1/2)c_\sigma(k_2/2)s_\sigma(2p-2k_1-k_2) \nn\\
  &&+8~\delta_{\mu,\sigma}~(1-\delta_{\mu,\rho})
  c_\sigma(p-k_1-k_2/2)c_\rho(k_1/2)s_\rho(2p-k_1)~. \nn
\end{eqnarray}

The integral symbols with subscripts for
fermion and gluon momenta, denoted by the letters $p$ and $k$
respectively, actually denote a sum over the momentum modes
on finite $N_\sigma^3\times N_\tau$. They are defined as:

\begin{eqnarray}
\int_k&=&{1\over N_\sigma^3N_\tau}\sum_k ~~{\rm with} \left\{
\begin{array}{l@{~,~}l}
  k_\mu={2\pi n_\mu\over N_\sigma} & -{1\over2}N_\sigma \le n_\mu\le {1\over
  2}N_\sigma-1,~\mu\not= 4\\
  k_4={2\pi n_4\over N_\tau} & -{1\over2}N_\tau\le n_4\le {1\over 2}N_\tau-1
\end{array}\right.,\nn\\
\int_p&=&{1\over N_\sigma^3N_\tau}\sum_p ~~{\rm with} \left\{
\begin{array}{l@{~,~}l}
  p_\mu={2\pi n_\mu\over N_\sigma} &  0\le n_\mu\le N_\sigma-1,~\mu\not= 4\\
  p_4={\pi\left( 2n_4+1\right)\over N_\tau} & 0\le n_4\le N_\tau-1
\end{array}\right..\nn
\end{eqnarray}
The shift of temporal fermion momentum modes, $p_4$, represents antiperiodic boundary
conditions in the time direction. As we consider the infinite volume limit the sum
becomes an integral over the Brillouin zone, $1/N_\sigma\sum_{l_i}
\rightarrow 1/(2\pi)\int_{-\pi}^\pi{\rm d}l_i$, $i=1,2,3$, for both fermionic
and gluonic momenta, $l=p,k$. The same holds for the time direction in the zero
temperature contributions.

The path integrals over the fermionic and gluonic fields are evaluated using
the identities
\begin{eqnarray}
  \left\langle A^a_\mu(k)A^b_\nu(-l)\right\rangle_0
  &=&{N_c^2-1 \over 2N_c}~{\Delta^{-1}_G}_{\mu,\nu}(k)~
  \delta_{k,l}~\delta^{a,b}~~,\nn\\
  \left\langle \bar{\psi}_\alpha^a(p)\psi^b_\beta(q)\right\rangle_0
  &=&{\left(
      i\sum_\mu\gamma_\mu h_\mu(p)-m
    \right)_{\alpha,\beta} \over s_F(p)}~
  \delta_{p,q}~\delta^{a,b}~~,\nn
\end{eqnarray}
with
\begin{eqnarray}
s_F(p)&=&\sum_\mu h_\mu^2(p)+m^2~~.\nn
\end{eqnarray}

\subsection{Tree level coefficients for improved rotational symmetry}
\label{app_2}
An expansion of the free inverse fermion propagator up to order $p^4$ yields
$$
D^{(0)}(p)=\sum_\mu h_\mu(p)h_\mu(p)=\sum_\mu A p_\mu^2 A
\left(
A+2B_1p_\mu^2+2B_2\sum_{\nu\not= \mu} p_\nu^2
\right)+\mathcal{O}(p^6)~~,
$$
with coefficients
\begin{eqnarray}
A&=&
  2c^{(0)}_{1,0}+12c^{(0)}_{1,2}+6c^{(0)}_{3,0}~~,\nn\\
B_1&=&
 -{1\over 3}c^{(0)}_{1,0}-2c^{(0)}_{1,2}-9c^{(0)}_{3,0}~~,\nn\\
B_2&=&
  -8c^{(0)}_{1,2}~~.\nn
\end{eqnarray}
Obviously the condition $B_1=B_2$ has to be satisfied to achieve rotational
symmetry up to order $p^4$, which leads to the constraint
\begin{eqnarray}
  c^{(0)}_{1,0}+27~c^{(0)}_{3,0}+6~c^{(0)}_{1,2}&=&24~c^{(0)}_{1,2}~~. \nn
\end{eqnarray}
The expansion shows that setting $c^{(0)}_{1,2}\equiv 0$, \ie the Naik-action,
leads to vanishing $\mathcal{O}(p^4)$ coefficients $B_1=B_2\equiv 0$, which
corresponds even to an $\mathcal{O}(a^2)$ improvement.

\subsection{Improved rotational symmetry at $\mathcal{O}(g^2)$}
\label{app_3}
The explicite expressions for the $\mathcal{O}(g^2)$ contributions to the
fermion propagator are:
\begin{eqnarray}
  D^{(0)}_\mu(p)&=&-h_\mu(p)~,\nn\\
  D^{(2)}_\mu(p)&=&-{N_c^2-1\over 2N_c}h^{(2)}_\mu(p)~,\nn\\
  \Sigma_\mu(p)&=&\mathcal{K}_\mu(p)+\mathcal{L}_\mu(p)~,\nn
\end{eqnarray}

\begin{eqnarray}
  \mathcal{K}_\mu(p)&=&{N_c^2-1\over 2N_c}\int_k{1\over s_F(p-k)}
  \sum_\nu\sum_{\rho,\bar{\rho}}\bigg[
    h_\mu(p-k) K_{\nu;\rho}(p,k)
    K_{\nu;\bar{\rho}}(p-k,-k)\nn\\
    &&\qquad\qquad\qquad\qquad -2h_\nu(p-k)K_{\mu;\rho}(p,k)
    K_{\nu;\bar{\rho}}(p-k,-k)
  \bigg]{\Delta_G^{-1}}_{\rho,\bar{\rho}}(k)~,
  \nn\\
  \mathcal{L}_\mu(p)&=&-{N_c^2-1\over 4N_c}\int_k
  \sum_{\rho,\bar{\rho}}L_{\mu;\rho,\bar{\rho}}(p,k,-k)
  {\Delta_G^{-1}}_{\rho,\bar{\rho}}(k)~.\nn
\end{eqnarray}
For $\mathcal{L}_\mu(p)$ which originates from the $S^{(2)}_F$ term
the $p$ dependence can be factored out. To achieve this one has to express the
trigonometrical functions of sums of $p$ and $k$ as sums of factorized
ones. Then taking into account that ${\Delta_G^{-1}}_{\mu,\mu}(k)$ is an even
function of all $k$-directions and that ${\Delta_G^{-1}}_{\mu,\nu}(k)$ is an
odd function of $k_\mu$ and $k_\nu$ and an even function of the remaining
$k$-directions one finds that a lot of terms are odd functions
in some $k$-direction and therefore vanish upon integration. The remaining
terms, without using fat links, \ie taking $\omega=0$,
include only three gluonic integrals:
\begin{eqnarray}
\mathcal{L}_\mu(p)&=&-{N_c^2-1\over 4N_c}\Bigg\{
s_\mu(p)~c_{1,0}^{(0)}~2~\mathcal{I}_1
+s_\mu(3p)~c_{3,0}^{(0)}~2~\mathcal{I}_2\nn\\
&&\qquad\qquad~+s_\mu(p)\sum_{\nu\not=\mu}c_\nu(2p)~~
c_{1,2}^{(0)}\left(4~\mathcal{I}_1+16~\mathcal{I}_3\right)\nn
\Bigg\}~~,
\end{eqnarray}
where the gluonic integrals are defined as
\begin{eqnarray}
\mathcal{I}_1&\equiv&\int_k{\Delta_G^{-1}}_{1,1}(k)~~,\nn\\
\mathcal{I}_2&\equiv&\int_k\left(
9-8s_1^2(k/2)-4s_1^2(k)
\right){\Delta_G^{-1}}_{1,1}(k)~~,\nn\\
\mathcal{I}_3&\equiv&\int_k\left(
  c_2^2(k/2){\Delta_G^{-1}}_{2,2}(k)
  -s_1(k/2)c_2(k/2)s_2(k){\Delta_G^{-1}}_{1,2}(k)
\right)~~.\nn
\end{eqnarray}
A calculation of these integrals to high accuracy using Gauss-Legendre
integration gives
\begin{eqnarray}
\mathcal{I}_1&=&0.1282908~~,\nn\\
\mathcal{I}_2&=&0.0752730~~,\nn\\
\mathcal{I}_3&=&0.5074740~~.\nn
\end{eqnarray}

\subsection{Contributions to free energy density}
\label{app_4}
The explicite expressions for the fermionic contributions to the free energy
density up to $\mathcal{O}(g^2)$ are:

\begin{eqnarray}
f_0&=&-2N_c\int_p\ln\left[s_F(p)\right]~,\nn\\
\left\langle
  \left(S_F^{(1)}\right)^2
\right\rangle_0&=&
2(N_c^2-1)\int_p\int_k{1\over s_F(p)s_F(p-k)}\sum_{\mu,\bar{\mu}}\Bigg\{
\nn\\
&&
\left[
  \delta_{\mu,\bar{\mu}}
  \left(
    \sum_\rho h_\rho(p)h_\rho(p-k)+m^2
  \right)
  -2h_\mu(p)h_{\bar{\mu}}(p-k)
\right]
\nn\\
&&\sum_{\rho,\bar{\rho}}K_{\mu;\rho}(p,k)K_{\bar{\mu};\bar{\rho}}(p-k,-k)
{\Delta_G^{-1}}_{\rho,\bar{\rho}}(k)
\Bigg\}~,
\nn\\
\left\langle
  S_F^{(2)}
\right\rangle_0&=&
(N_c^2-1)\int_p\int_k{1\over s_F(p)}\sum_{\mu}
h_\mu(p)
\sum_{\rho,\bar{\rho}}L_{\mu;\rho,\bar{\rho}}(p,k,-k)
{\Delta_G^{-1}}_{\rho,\bar{\rho}}(k)~,\nn\\
\left\langle
  \tilde{S}_F^{(0)}
\right\rangle_0&=&-(N_c^2-1)\int_p\sum_\mu {h_\mu(p)h_\mu^{(2)}(p)\over s_F(p)}
~.
\nn
\end{eqnarray}
In $\left\langle S_F^{(2)} \right\rangle_0$ the $k$ integration/sum can
again be factored, similarly as for $\mathcal{L}_\mu(p)$ in \ref{app_3}.
\end{appendix}

\vfill\eject

\end{document}